\documentclass[pre,twocolumn,showpacs,preprintnumbers,amsmath,amssymb]{revtex4-1}

\pdfoutput=1 

\usepackage{graphicx}
\usepackage{dcolumn}
\usepackage{bm}

\begin{document}

\newcommand{\bea}{\begin{eqnarray}}
\newcommand{\eea}{  \end{eqnarray}}
\newcommand{\bit}{\begin{itemize}}
\newcommand{\eit}{  \end{itemize}}

\newcommand{\be}{\begin{equation}}
\newcommand{\ee}{\end{equation}}
\newcommand{\ra}{\rangle}
\newcommand{\la}{\langle}
\newcommand{\U}{\widetilde{U}}


\def\bra#1{{\langle#1|}}
\def\ket#1{{|#1\rangle}}
\def\bracket#1#2{{\langle#1|#2\rangle}}
\def\inner#1#2{{\langle#1|#2\rangle}}
\def\expect#1{{\langle#1\rangle}}
\def\e{{\rm e}}
\def\proj{{\hat{\cal P}}}
\def\tr{{\rm Tr}}
\def\H{{\hat H}}
\def\Hdag{{\hat H}^\dagger}
\def\Lop{{\cal L}}
\def\Ehat{{\hat E}}
\def\Edag{{\hat E}^\dagger}
\def\Shat{\hat{S}}
\def\Sdag{{\hat S}^\dagger}
\def\Ahat{{\hat A}}
\def\Adag{{\hat A}^\dagger}
\def\U{{\hat U}}
\def\Udag{{\hat U}^\dagger}
\def\Zhat{{\hat Z}}
\def\Phat{{\hat P}}
\def\Op{{\hat O}}
\def\id{{\hat I}}
\def\x{{\hat x}}
\def\P{{\hat P}}
\def\Px{\proj_x}
\def\Pr{\proj_{R}}
\def\Pl{\proj_{L}}


\title{Correspondence behavior of classical and quantum dissipative directed transport via thermal noise}

\author{Gabriel G. Carlo}
\affiliation{Departamento de F\'\i sica, CNEA, Libertador 8250,
(C1429BNP) Buenos Aires, Argentina}
\author{Leonardo Ermann}
\affiliation{Departamento de F\'\i sica, CNEA, Libertador 8250,
(C1429BNP) Buenos Aires, Argentina}
\author{Alejandro M. F. Rivas}
\affiliation{Departamento de F\'\i sica, CNEA, Libertador 8250,
(C1429BNP) Buenos Aires, Argentina}
\author{Mar\'\i a E. Spina}
\affiliation{Departamento de F\'\i sica, CNEA, Libertador 8250,
(C1429BNP) Buenos Aires, Argentina}

\email{carlo@tandar.cnea.gov.ar,ermann@tandar.cnea.gov.ar,rivas@tandar.cnea.gov.ar,spina@tandar.cnea.gov.ar}

\date{\today}


\begin{abstract}

We systematically study several classical-quantum correspondence properties of the dissipative modified kicked rotator, 
a paradigmatic ratchet model. 
We explore the behavior of the asymptotic currents for finite $\hbar_{\rm eff}$ values in a 
wide range of the parameter space. We find that the correspondence between the classical currents 
with thermal noise providing fluctuations of size $\hbar_{\rm eff}$ and the quantum ones without it, is very good in general 
with the exception of specific regions. 
We systematically consider the spectra of the corresponding classical Perron-Frobenius operators and quantum superoperators. 
By means of an average distance between the classical and quantum sets of eigenvalues 
we find that the correspondence is unexpectedly quite uniform. This apparent contradiction is 
solved with the help of the Weyl-Wigner distributions of the equilibrium eigenvectors, which reveal the key 
role of quantum effects by showing surviving coherences in the asymptotic states. 

\end{abstract}

\maketitle

\section{Introduction}
\label{sec1}

A pioneering paper by Feynman \cite{Feynman} re-ignited the interest in directed transport 
\cite{Reimann,Kohler,Denisov}, this meaning unbiased transport phenomena in 
systems which are driven out of equilibrium. There is a wealth of fields for their 
application such as biology \cite{biology}, nanotechnology \cite{nanodevices}, chemistry \cite{chemistry}, 
cold atoms \cite{CAexp,AOKR}, and Bose-Einstein condensates \cite{BECratchets,purelyQR,QR2,coherentControl}, 
for example. Among all these alternatives, we focus in deterministic ratchets with dissipation 
generally associated with a classical asymmetric chaotic attractor \cite{origin,Mateos}; the quantum versions lead 
to interesting applications in cold atoms \cite{qdisratchets}. The classical aspects of the 
parameter space of this system have been studied in detail in \cite{Celestino}. 
These new results revealed that the isoperiodic stable structures (ISSs, Lyapunov stable islands) 
have a fundamental role in the current shape. The quantum counterparts of these structures 
(QISSs) \cite{Carlo,Ermann} have proven to be very well approximated by means of a 
thermal coarse graining of the classical dynamical equations
(i.e., adding thermal noise of the order of $\hbar_{\rm eff}$) 
in representative cases. Taking into account these examples, it was recently found \cite{Carlo2} 
that the Perron-Frobenius operators associated with the classical evolution
with thermal noise and the quantum superoperators without it show very similar spectra. 
It deserves noticing that the study of Perron-Frobenius operators in the Ulam approximation 
is very useful in open systems theory \cite{Kullig}

In a recent publication a semiclassical approach 
has been taken into account \cite{Beims}. There, it was argued that an effective (semi)classical map with 
noise could be used as a direct replacement for the quantum system itself and several consequences 
have been derived. We take into account a 
thermal-like noise, i.e. a Gaussian noise whose strength is given by $\hbar_{\rm eff}$, leaving 
no free parameters. 
It is important to underline that the effective temperature to which it can be associated is different throughout the 
parameter space. Nevertheless, both our thermal noise approach and the semiclassical one 
are very much alike. In this work we verify our conjecture about the role of the coarse graining due to 
quantization: it shares the main properties of a thermal coarse graining for this kind of dissipative 
systems. In fact, it induces chaotic behaviour in situations where one expects a simple attractor for instance \cite{Carlo}. 
Thus, it is of general nature and is applicable to any finite $\hbar_{\rm eff}$ value, 
including the semiclassical regime which of course is compatible with it. 
Finally, the quantum regime is not only interesting from a theoretical point of view, but also for 
a great number of experimental situations in which the semiclassical limit is not reached.

However, there are relevant finite $\hbar_{\rm eff}$ quantum effects that cannot be reproduced by a classical map. 
In this paper we find that, though the general correspondence is very good, there are specific 
regions in which this mechanism shows limitations. This is clearly seen with the help of the main quantity of 
interest in directed transport, i.e. the asymptotic current, keeping in mind that it is 
just an average quantity that does not reflect a complete picture. 
Also, it is important to notice that transitory regimes are interesting but they are not the 
main objective here. But, when we study the spectra of the Perron-Frobenius operators with thermal 
noise and the quantum superoperators without it we find an almost uniform correspondence, 
with no clear signs of these discrepancy regions. This could be puzzling in view of the current behaviour previously 
mentioned. Moreover, we already know \cite{Carlo} that chaotic limit sets provide a quite 
deal of mixing with no need of noise, making uniformity unexpected. On the other hand, 
even for the chaotic case the thermal coarse graining is necessary to smooth out the classical fluctuations 
not present in the quantum version and in order to make both spectra agree \cite{Carlo2}.
To clarify this we explore the morphology of the eigenvectors associated with the equilibrium eigenvalues. 
Here we find that surviving coherences (that cannot be reproduced by means of a classical model with 
noise) are a key indicator to understand the limits that this mechanism shows in describing some asymptotic 
currents. This also provides valuable 
information in order to develop a semiclassical approximation of the equilibrium eigenstates -- 
a long standing objective -- perhaps taking into 
account convenient wavepackets and the classical dynamics with thermal noise.

This paper has the following structure: In Sec. \ref{sec2} we introduce our model 
which is a modified kicked rotator with dissipation, clearly explaining the way 
in which we add the thermal noise in the classical model to find the 
correspondence with the quantum one. In Sec. \ref{sec3} we explore the difference 
between the classical and quantum asymptotic currents in the parameter space finding 
that there are regions in which the correspondence is not so accurate. 
In Sec. \ref{sec4} we systematically explore the spectra of the classical Perron-Frobenius operator 
with thermal noise and the quantum superoperator in the parameter space, which show a seemingly uniform 
correspondence. In Sec. \ref{sec5} we explain this apparent contradiction by extending this 
study to the phase space and using the Weyl-Wigner distributions of the equilibrium eigenvectors. 
In Sec. \ref{sec6}, we present our conclusions.

\section{Model and calculation methods}
\label{sec2}

We consider a particle moving in one dimension
[$x\in(-\infty,+\infty)$] periodically kicked by the asymmetric potential:
\begin{equation}
V(x,t)=k\left[\cos(x)+\frac{a}{2}\cos(2x+\phi)\right]
\sum_{m=-\infty}^{+\infty}\delta(t-m \tau),
\end{equation}
where $k$ is the strength of each kick and $\tau$ is the kicking period. 
When adding dissipation we obtain a dissipative ratchet system that can be 
written as the following map \cite{qdisratchets,Celestino}
\begin{equation}
\left\{
\begin{array}{l}
\overline{n}=\gamma n +
k[\sin(x)+a\sin(2x+\phi)]
\\
\overline{x}=x+ \tau \overline{n}.
\end{array}
\right.
\label{dissmap}
\end{equation}
Here $n$ is the momentum variable conjugated to $x$ 
and $\gamma$ ($0\le \gamma \le 1$) is the dissipation parameter.
The conservative limit is reached at $\gamma=1$, whereas the 
value $\gamma=0$ gives the maximum damping. In order to simplify 
the parametric dependence it is usual to introduce a rescaled 
momentum variable $p=\tau n$ and the quantity $K=k \tau$. 
The directed current emerges as a consequence of breaking the 
spatial and temporal symmetries by adopting $a \neq 0$ with $\phi \neq m
\pi$, and  $\gamma \neq 1$, respectively. It is worth mentioning that we take $a=0.5$ 
and $\phi=\pi/2$ throughout this work.

We have conjectured \cite{Carlo} that the main effects of the quantum 
fluctuations are similar to those of Gaussian fluctuations of the order of $\hbar_{\rm eff}$ 
in the classical analogue ($\hbar_{\rm eff}$ is 
the effective Planck constant to be defined in the next paragraph). 
In order to introduce it we take $\overline{n}' \rightarrow
\overline{n}$ in Eq. \ref{dissmap}, where
$\overline{n}'=\overline{n} + \xi$. Though the essential idea is the Gaussian 
nature of fluctuations on the effective Planck scale, we can associate the noise
variable $\xi$ with a temperature $T$ by means of the relation $
<\xi^2> =2 (1-\gamma) k_B T$, where $k_B$ is the Boltzmann
constant (which we take equal to 1). Moreover, in this work we take $T
= \hbar_{\rm eff}/[2 (1-\gamma)]$. Note that we have explicitly fixed the value of $T$ 
as a function of $\hbar_{\rm eff}$ and $\gamma$. This leaves no free parameters 
in order to test the behaviour of our conjecture in this situation. However 
we underline that this is not essential for it to be valid. In the following, 
when we refer to classical properties or quantities it is assumed that they 
correspond to the classical system with thermal noise unless otherwise stated.

The corresponding quantum model without thermal noise is given by:
$x\to \hat{x}$, $n\to \hat{n}=-i (d/dx)$ ($\hbar=1$).
Since $[\hat{x},\hat{p}]=i \tau$ (where $\hat{p}=\tau \hat{n}$), the effective Planck constant
is $\hbar_{\rm eff}=\tau$. The classical limit corresponds to
$\hbar_{\rm eff}\to 0$, while $K=\hbar_{\rm eff} k$ remains constant. 
We fix $\hbar_{\rm eff}=0.137$ in this work since we are not interested in reaching 
the classical limit. 
Dissipation at the quantum level is introduced by means of the
master equation \cite{Lindblad} for the density operator $\hat{\rho}$ of the
system
\begin{equation}
\dot{\hat{\rho}} = -i
[\hat{H}_s,\hat{\rho}] - \frac{1}{2} \sum_{\mu=1}^2
\{\hat{L}_{\mu}^{\dag} \hat{L}_{\mu},\hat{\rho}\}+
\sum_{\mu=1}^2 \hat{L}_{\mu} \hat{\rho} \hat{L}_{\mu}^{\dag} \equiv \Lambda \rho.
\label{lindblad}
\end{equation}
Here $\hat{H}_s=\hat{n}^2/2+V(\hat{x},t)$ is the system
Hamiltonian, \{\,,\,\} is the anticommutator, and $\hat{L}_{\mu}$ are the Lindblad operators
given by \cite{Dittrich, Graham}
\begin{equation}
\begin{array}{l}
\hat{L}_1 = g \sum_n \sqrt{n+1} \; |n \rangle \, \langle n+1|,\\
\hat{L}_2 = g \sum_n \sqrt{n+1} \; |-n \rangle \, \langle -n-1|,
\end{array}
\end{equation}
with $n=0,1,...$ and $g=\sqrt{-\ln \gamma}$ (due to the Ehrenfest theorem).

All this is enough to compare the behaviour of the asymptotic currents. 
But if we want to go further in a systematic study of the classical 
to quantum correspondence we need to incorporate more quantities. 
The classical densities in phase space evolve with the
Perron-Frobenius operator arising from the Liouville equation
corresponding to the map in Eq. \ref{dissmap}. A discretization of phase space gives rise 
to the Ulam method \cite{Ulam}, which is a coarse grained approximation to
the Perron-Frobenius operator \cite{UlamDMaps}. The procedure consists of defining the 
Ulam matrix $S$ by means of dividing the phase space into $M^2$ cells and
propagating $n_{\rm tr}$ random points from each cell $j$ with the classical map. 
The number $n_{ij}$ of trajectories
arriving to cell $i$ from the cell $j$ allows to write the elements of $S$ as 
$S_{ij} = {n_{ij} \over n_{\rm tr}}$. It is important to notice that this discretization
is comparable to a diffusive noise of order $h_{\rm eff}^{\rm
PF} \propto {1 \over M}$ . For homogeneous systems and
sufficiently large values of $M$ the Ulam method converges to the spectrum of 
the continuous system. If the coarse graining of the Ulam method is smaller than that 
attributable to the thermal fluctuations (i.e., $h_{\rm eff}^{\rm PF} \le h_{\rm eff}$) 
the results obtained are independent of $h_{\rm eff}^{\rm PF}$. 
In the following the expression Perron-Frobenius operator will make reference 
to its Ulam approximation. 

In the quantum case the evolution of the density matrix is given by $\rho_{t+1} = e^ {\Lambda} \rho_{t}$,
where $e^ {\Lambda}$ is a non-unital superoperator of dimension $N^2 \times N^2$ constructed
by numerical integration of Eq. \ref{lindblad}. Here $h_{\rm eff} \propto {1 \over N}$.
For the diagonalization of $S$ and $e^
{\Lambda}$ we have used the Arnoldi method.

\section{Classical and quantum asymptotic currents}
\label{sec3}

\begin{figure}[htp]
\includegraphics[width=0.47\textwidth]{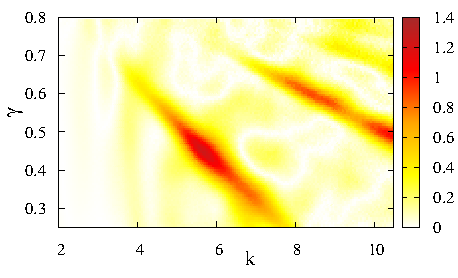} 
\includegraphics[width=0.47\textwidth]{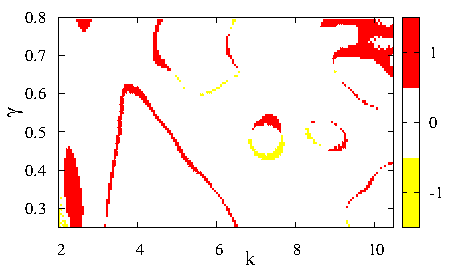} 
 \caption{(color online) In the top panel we display the  distance between the classical and quantum 
 asymptotic currents given by $|J_{\rm c}^{\rm th}-J_{\rm q}|$ . We have considered this in order to better 
 reflect the discrepancy in the approximation. In the bottom panel we show 
 the sign difference $({\rm sgn}(J_{\rm c}^{\rm th})-{\rm sgn}(J_{\rm q}))/2$.
}
 \label{fig1}
\end{figure}

We first look at the properties of asymptotic ratchet currents 
$J=<p>$ (where $<>$ stands for either the classical or quantum averages). A word of caution is in 
order here; this quantity is an average and as such the information it provides regarding classical-quantum 
correspondence is limited. Nevertheless it is the main quantity of interest for directed transport systems, along 
with the  widths of distributions \cite{Ermann}. In the top panel of Fig. \ref{fig1} 
we show the difference $|J_{\rm c}^{\rm th}-J_{\rm q}|$, 
where $J_{\rm c}^{\rm th}$ stands for the classical current and 
$J_{\rm q}$ for the quantum one. This shows the distance between them, i.e. the discrepancy in the approximation. 
The differences are in general quite small as can be seen from the white-yellow overall appearance of the plot.
For this statement to be meaningful one has to keep in mind the much higher currents of the original 
classical system without thermal noise. Also, one should remember that the width in $p$ of the distributions can be two orders of 
magnitude greater than these values. This is a proof that, regarding currents, our conjecture works 
very well, despite we have fixed the size of the fluctuations at exactly $\hbar_{\rm eff}$, 
a too restrictive requirement. 

But this image also shows that a plain replacement of the quantum currents with 
this kind of approximations could lead to quantitatively wrong results. This is specially evident in some 
regions of the isoperiodic structures $B_1$ and $B_2$ \cite{Beims}. There are also minor differences in the other 
$B$ structures and in a few cases of chaotic regions. 
It is also interesting to note that some of the better correspondence arises in borders of what is left of the 
original classical parameter space regions associated with either chaotic or regular behaviour. These are small fluctuations 
but still give a hint on a missing ingredient that could be responsible for the enhancement of one or the other structure at 
either side of this frontier. In the following sections we will explain its nature.

In the bottom panel of Fig. \ref{fig1} the sign difference 
$({\rm sgn}(J_{\rm c}^{\rm th})-{\rm sgn}(J_{\rm q}))/2$ is displayed. This shows 
that for almost all cases the sign is very well reproduced, but also that -- though being quite similar -- 
the borders of the positive and negative regions do not necessarily coincide.

\section{Spectral correspondence}
\label{sec4}

\begin{figure}
  \includegraphics[width=0.47\textwidth]{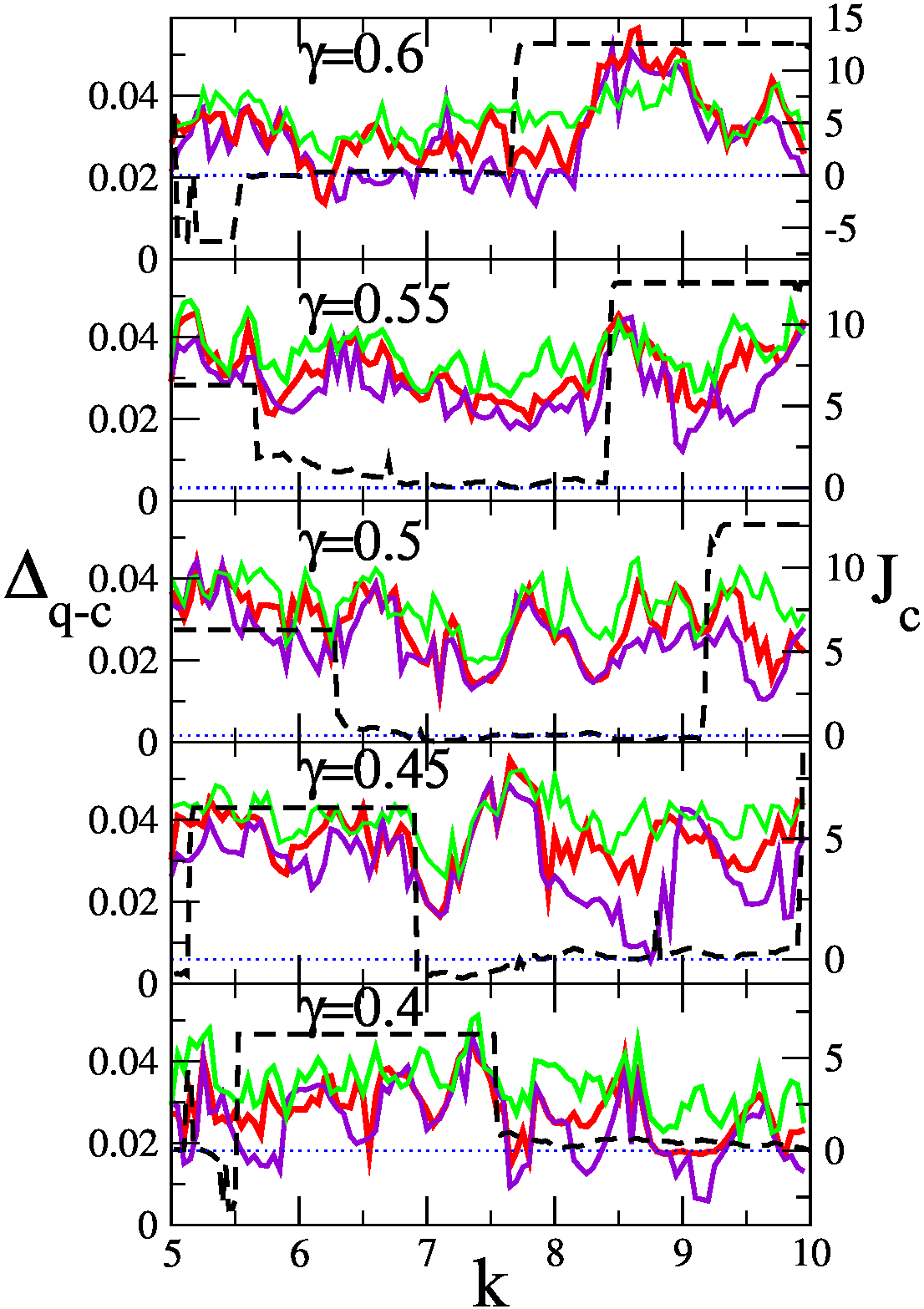} 
 \caption{(color online) Average distance between classical and quantum eigenvalues $\lambda$ in 
 complex plane vs. $k$ for values of $\gamma=0.4;0.45;0.5;0.55;0.6$ from the bottom to the top panel, 
 correspondingly. Green, red and violet solid lines show the average distance for eigenvalues with modulus 
 larger than $0.25$, $0.35$ and $0.4$ respectively (values are shown in left axis).
 Dashed black lines represent classical currents (with values shown in right axis) with dotted blue lines
 showing zero current.
 }
 \label{fig2}
\end{figure}

The asymptotic current is the main quantity of interest in directed transport but, being an 
average is not a very precise gauge to measure the correspondence behavior which we are interested 
to test in this work. Then we turn to study complementary quantities. In this Section 
we look at the properties of the spectra of the Perron-Frobenius operators and 
the quantum superoperators \cite{Carlo2}. For that we define the average distance 
between the classical and the quantum spectra $\Delta_{\rm q-c}$ as the average of the euclidean distances in the complex plane 
between the eigenvalues $\lambda$ that are nearest neighbours from both spectra. In Fig. \ref{fig2} we show this measure 
for five different values of the dissipation parameter, $\gamma=0.4;0.45;0.5;0.55;0.6$ 
(from the bottom panel to the top panel, correspondingly), 
as a function of $k$. We have considered just the longest-lived eigenvalues, but in order to 
check the convergence of this measure we have taken into account three different values for 
the minimum modulus required to be considered as such ($0.25$, $0.35$ and $0.4$, by means of 
green, red and violet solid lines, respectively). The 
behaviour is quite consistent and these curves show no meaningful jumps, being all of them 
around one order of magnitude lower than the minimum moduli. This not only tells us that the 
spectral correspondence is good but that it is approximately uniform. As a guide to the eye 
we have also drawn the classical $J_{\rm c}$ without noise, so as to reflect the insensitivity 
of this measure to the changes in the nature of the dynamics that takes place in the 
parameter space. This is an unexpected result a priori, since in Sec. \ref{sec3} it became 
clear that the approximation does not work in a uniform way throughout the parameter space. 
Moreover, we already knew from our previous work that the chaotic currents are already 
in good correspondence without needing extra fluctuations. This guarantees non uniformity, 
no matter which noise model one chooses.

\begin{figure}
  \includegraphics[width=0.47\textwidth]{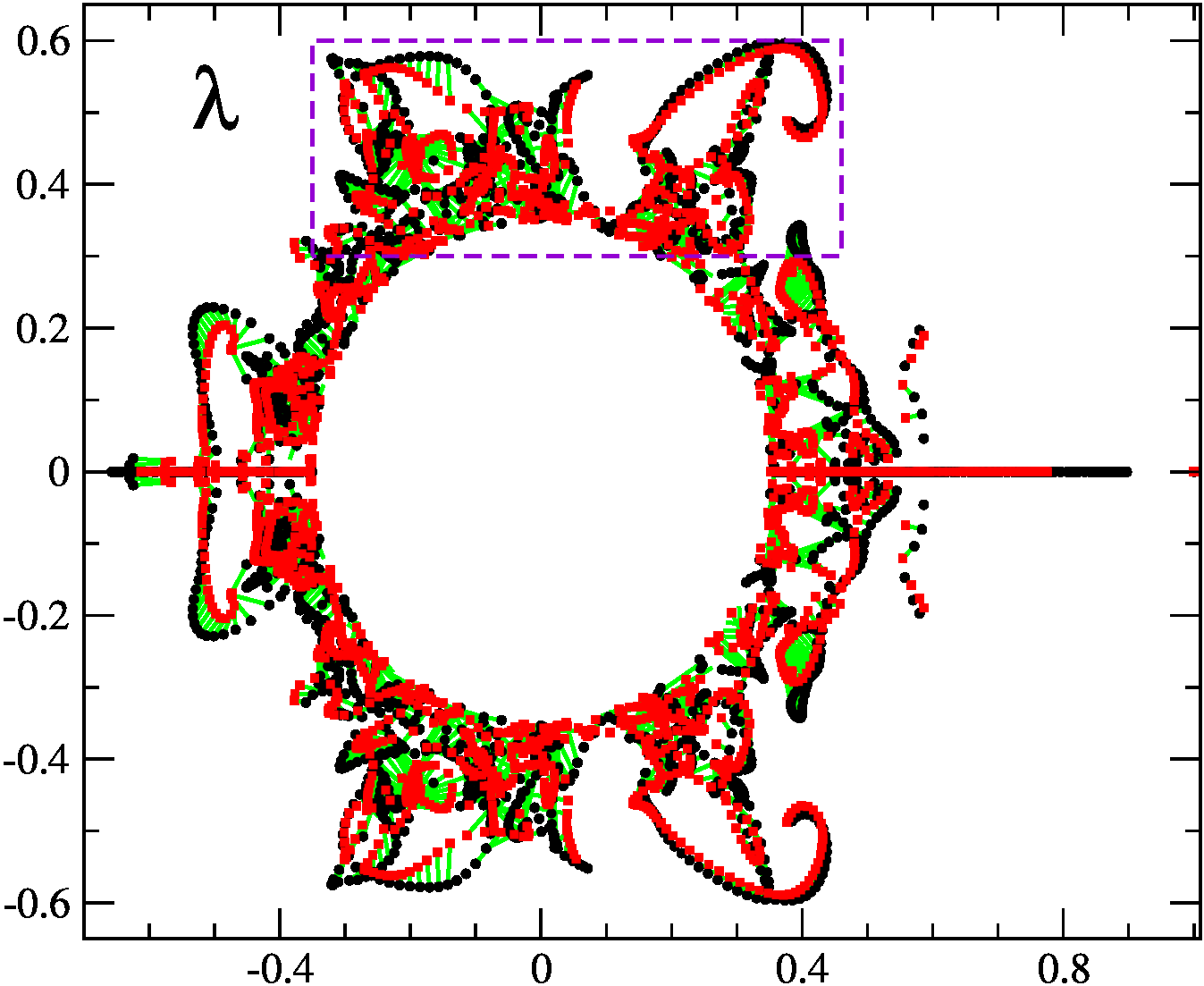}\\
  \includegraphics[width=0.47\textwidth]{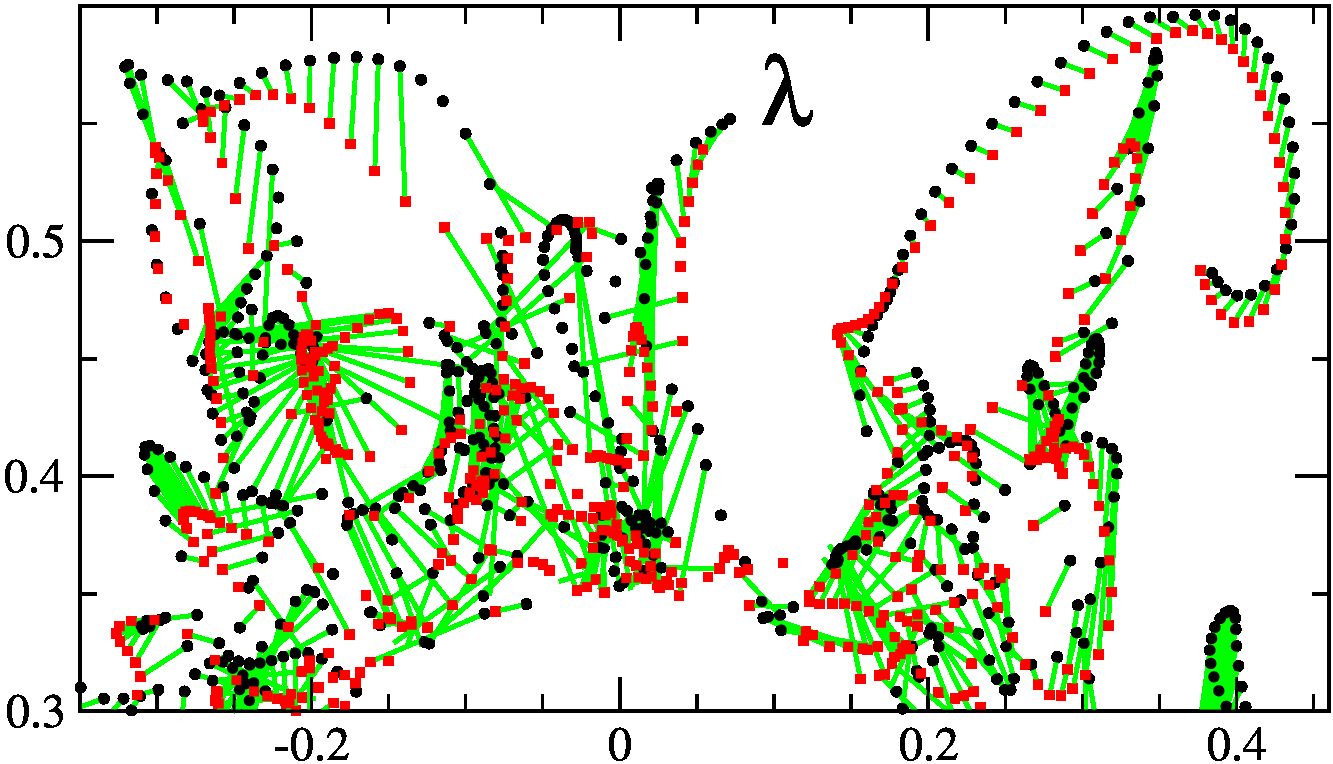} 
 \caption{(color online) Evolution of distances between classical and quantum eigenvalues $\lambda$ in the 
complex plane for $\gamma=0.6$ and $k$ between 5 to 10 (with steps $\Delta k=0.05$. 
 Black circles and red squares show the eigenvalues of classical and quantum systems respectively, 
 while green lines represent the distance between both of them. Distance is only computed when classical 
 and quantum eigenvalues are larger than $0.35$.
 The highlighted area (with a dashed violet line) of the top panel is shown in the bottom panel.
 }
 \label{fig3}
\end{figure}

When we look deeper into the details of this correspondence we see how the spectra follow each 
other as the parameter $k$ is varied. In the top panel of Fig. \ref{fig3} we show the distances between the 
classical and quantum eigenvalues for $\gamma=0.6$ and $k$ between $5$ to $10$, 
taking steps $\Delta k=0.05$. In the bottom panel of Fig. \ref{fig3} we display a zoom that highlights this remarkably 
uniform coincidence.

\section{Phase space representation of the asymptotic eigenstates}
\label{sec5}

What could explain this apparent paradox?  
We know from a previous work \cite{Carlo2} that even for the already similarly behaved 
chaotic cases one needs to smooth out classical fluctuations in order to recover 
a quantum-like spectrum. This would put on an equal footing the spectra for different 
regions of parameter space in terms of their classical-quantum correspondence. And one 
has to keep in mind that all the information needed to obtain the asymptotic currents 
is contained in the equilibrium eigenstates, which of course have all the same eigenvalue $1$. 
So, differences in these states could entirely explain the differences that we have found in 
Sec. \ref{sec3}, differences which are not necessarily reflected in the rest of the eigenvalues 
as a whole. 

On one hand we have the right invariant 
eigenvectors of the Perron-Frobenius operator $R_{\lambda_0}$ which are real and
non negative and can be directly taken as probability distributions in phase space. 
On the other, we have the Weyl-Wigner symbols for the right invariant eigenvectors of the
quantum superoperator $\hat{R}_{\lambda_0}$ which can be taken as density matrices 
satisfying $Tr(\hat{R}_{\lambda_0})=1$.
The Weyl-Wigner symbols for a $N$ dimensional Hilbert space are defined in a redundant $2N\times2N$
discrete phase space \cite{opetor}. This is formed by the grid of
points $x=\frac{1}{N}(a,b)$ with $a$ and $b$ semi integer numbers
running from $0$ to $N-1/2$. In this way, the Weyl-Wigner symbol
$R(x)$ of the operator $\hat{R}$ is obtained from its
 matrix elements in the coordinate representation as
\[
R(x)=\sum_{n=0}^{N-1}<q_{2b-n}|\hat{R}|q_{n}>\exp\left(\frac{i2\pi}{N}2a(b-n)\right).
\]
In order to get rid of redundancies and ``ghost images'' derived essentially from
the cylindrical topology of our phase space, we use a method that
has been developed by Arg\"uelles and Dittrich \cite{Ditt} (for more details on this calculation 
we refer to \cite{Carlo2}).

A convenient measure to compare both distributions is the overlap. 
To calculate these overlaps we take into account that
any state $\hat{R}$ can be represented by $R(x)$ with $x=(p,q)$ a
point in phase space. For the classical states, $R(x)$ stands for
the right eigenvector, while for the quantum ones, $R(x)$ is the
Weyl-Wigner symbol. Hence, given any two states $\hat{R_{1}}$ and
$\hat{R_{2}}$ , their overlap is defined as:
\[
O(\hat{R_{1}},\hat{R_{2}})=Tr\left(\hat{R_{1}}\hat{R_{2}}\right)/\sqrt{
\left[Tr\left(\hat{R_{1}^{2}}\right)Tr\left(\hat{R_{2}^{2}}\right)\right]}=
\]
\[
\sum_{x}R_{1}(x)R^{*}_{2}(x)/\sqrt{\left[\left(\sum_{x}|R_{1}(x)|^{2}\right)
\left(\sum_{x}|R_{2}(x)|^{2}\right)\right]},
\]
where $R^{*}(x)$ and $|R(x)|$ stand respectively for the complex
conjugate and absolute value of $R(x)$. The overlap defined above
is a complex magnitude, its modulus is invariant even though its
argument depends on the relative phase between the eigenvectors.
Also, when this relative phase is null $O(\hat{R_{1}},\hat{R_{2}})$ is real.

\begin{figure}
  \includegraphics[width=0.47\textwidth]{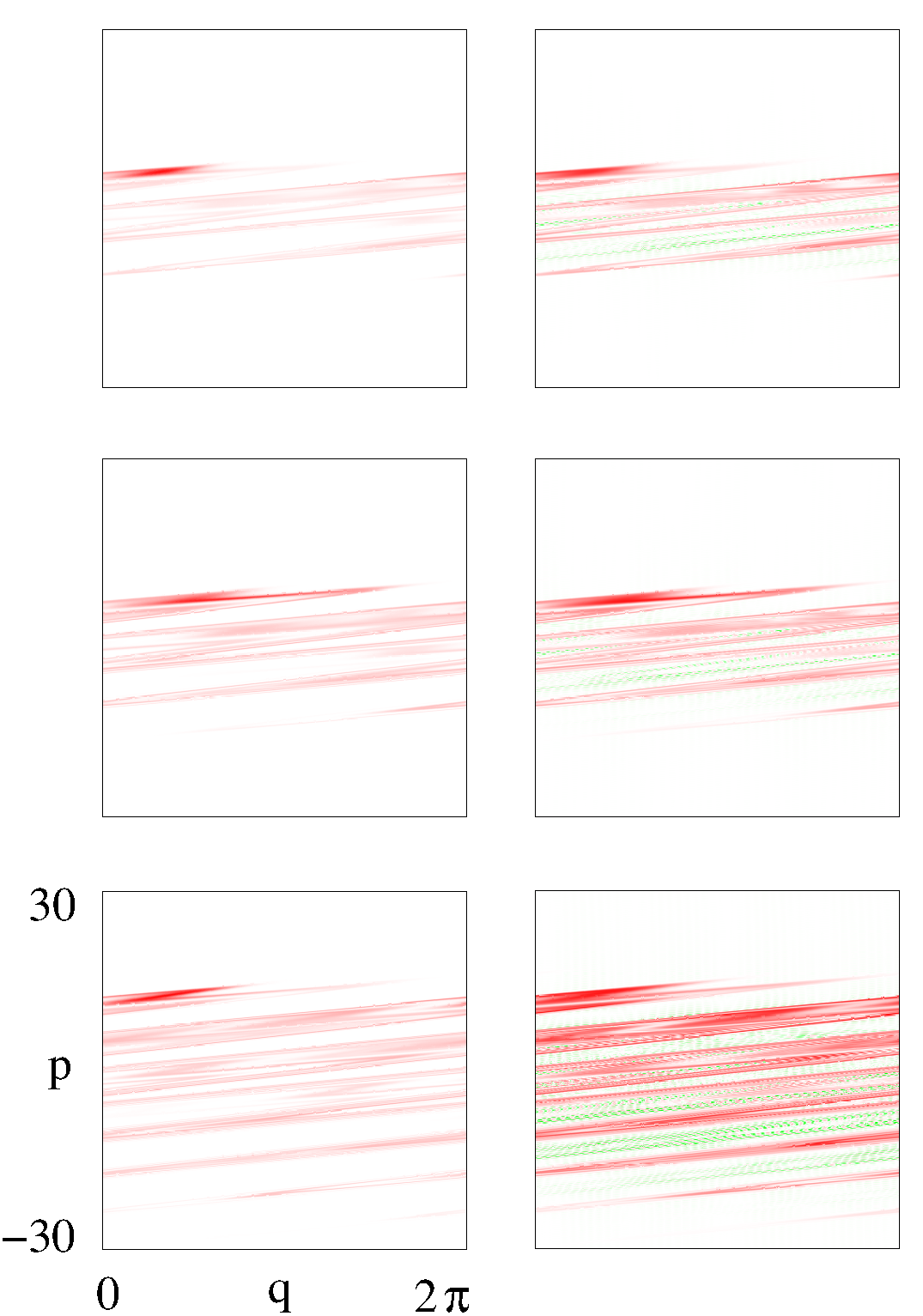} 
 \caption{(color online) Three examples of distributions of classical (left column) and 
 quantum equilibrium states (right column). Upper panels correspond to 
 $k=5$ and $\gamma=0.5$, middle panels to $k=5.55$ and $\gamma=0.55$, 
 and bottom panels to $k=9.25$ and $\gamma=0.55$. Negative values are highlighted in green 
 (light gray) while red (darker shades of gray) stand for the positive ones. 
 }
 \label{fig4}
\end{figure}
\begin{figure}
  \includegraphics[width=0.47\textwidth]{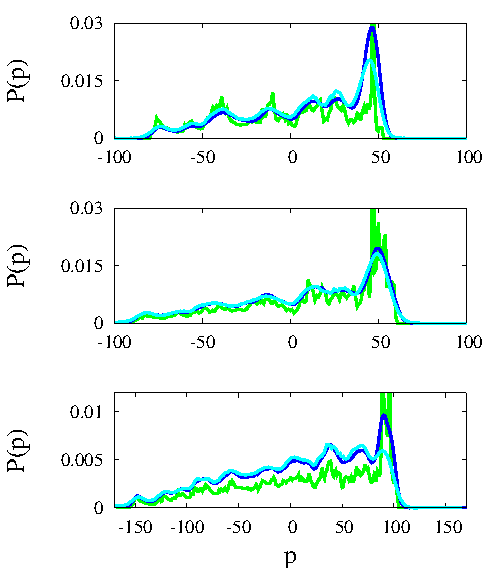} 
 \caption{(color online) $P(p)$ probability distributions as a function of $p$ 
 for the three cases shown in Fig. \ref{fig4}, in the same order. Blue (black) lines correspond to the 
 classical distributions with thermal noise, green (gray) lines to the ones without thermal noise, and 
 cyan (light gray) lines to the quantum distributions. 
 }
 \label{fig5}
\end{figure}
We will focus our attention on the greatest discrepancy regions $B_1$ and $B_2$, which explain 
almost all the main classical-quantum differences. 
In Fig. \ref{fig4} we show three examples, two of them where the discrepancy in the classical approximation 
is maximal (of the order of $1$ in the current), and one in which the performance is good. 
In the left column we display the classical distributions, and in the right one the quantum distributions.
In Fig. \ref{fig5} we show the probability distributions $P(p)$ as a function of $p$, that arise from the classical and 
quantum evolutions, in the same order as in the previous figure. 

The first example with $k=5$ and $\gamma=0.5$ belongs to a $B_1$ isoperiodic structure that is located 
in the red area of the largest dark region in the top panel of Fig. \ref{fig1}, which has the greatest differences 
in the current values (the maximum being around $1.3$, and in this case being of $0.74$). 
The overlap between both phase space distributions is $O=0.88$, which is good but if we look at the top panel 
of Fig. \ref{fig5} we can see that there is an excess of classical probability around the location of the limit 
cycle associated to this structure (see the green line corresponding to the classical distribution without noise). 
This suggests that there should be a probability rebalancing 
due to purely quantum effects that takes it from this peak and transfers it to the chaotic basin 
which is next to it. 

The second example with $k=5.55$ and $\gamma=0.55$ also belongs to $B_1$ but in this case it is located 
in a yellow area of this region in the top panel of Fig. \ref{fig1}, reflecting the fact that the current behaviour is quite 
similar. The overlap $O=0.94$ agrees with the difference in $J$ being just $0.13$, and the $P(p)$ distributions 
are almost the same (middle panel of of Fig. \ref{fig5}). 
If we compare both situations the main difference is the location of them in parameter space: while the first case 
is in what we would call the bulk of $B_1$ region the second is clearly at its border. This is the largest 
regular region and the influence of the surrounding areas of (mainly) chaos is reduced in its bulk. 
It seems that quantum effects enhance the probability transfer between these otherwise separated structures, 
going further than the rebalancing introduced by a plain coarse graining due to the size of $\hbar_{\rm eff}$. 
As a matter of fact, we can clearly see the quantum nature of the distributions by means of the interference 
fringes that reveal the persistence of coherences despite the interaction with a quantum dissipative environment 
(at least at this finite values of the effective Planck constant). 

Finally, the third example with $k=9.25$ and $\gamma=0.55$ belongs to the $B_2$ structure and is located in 
the second largest dark region in the top panel of Fig. \ref{fig1}. The overlap is now $O=0.86$ which represents 
a performance comparable to the first case, with a difference in the current values of $0.73$. 
When looking at the bottom panel of Fig. \ref{fig5}, the $P(p)$ distributions 
are also of the same nature than those of the first example. 
The behaviour is exactly the same as for $B_1$, so our explanation of these discrepancies 
is of general nature.

\section{Conclusions}
\label{sec6}

We have systematically studied the classical-quantum correspondence via thermal noise in a paradigmatic directed transport 
system, the dissipative modified kicked rotator. We have explored a good portion of the parameter 
space in terms of the asymptotic currents and the spectra of both, the classical operators and 
the quantum superoperators. The first approach, though confirming the overall correspondence gave us 
the expected picture of a non uniform shape, dependent on the kind of dynamics underlying each region. On the contrary, 
the spectral study resulted in a uniform behaviour. By focusing on the morphology of the equilibrium eigenstates, 
we could verify through well developed interference fringes, that the quantum nature is persistent at 
this finite $\hbar_{\rm eff}$, despite the environmental effects. This resilient quantumness is responsible 
for an enhancement of probability transfer from the main regular structures to the ones surrounding them 
in parameter space.

An effective (semi)classical map with noise has been proposed as a direct replacement 
for the quantum system and some results have been derived from this identification \cite{Beims}. 
This map is quite similar to our much simpler approach, which is just a map with thermal 
noise with no free parameters. Ours is based on the general assumption that the main features of the quantum coarse 
graining are similar to those of a Gaussian coarse graining with size of the order of $\hbar_{\rm eff}$, 
in this directed transport system. This can be considered simplistic but it turns 
out to be precise enough in the overwhelming majority of the cases. 

Of course, there are known discrepancies 
\cite{Carlo,Carlo2}. Again, we ascribe the difference in the asymptotic currents 
for some specific cases to finite $\hbar_{\rm eff}$ quantum effects. 
In \cite{ozoriobrodier} it has been shown that for dissipative
Markovian systems in the semiclassical limit, the Wigner function evolves
into a positive-definite phase-space distribution.
However, in Fig. \ref{fig4} the Wigner functions of the equilibrium eigenstates 
display  negative  values  due  to interference fringes.
This shows that for the value of $\hbar_{\rm eff}$ used here, we have not reached the
semiclassical limit. As such, we conclude that though 
the classical dynamics with Gaussian/thermal noise is a very good approximation to the quantum dynamics, 
quantum effects could still be non-neglectable, making 
a direct replacement an oversimplification. In fact, if one wants to develop a semiclassical 
approximation of theoretical interest and amenable to the vast majority of experimental 
situations in the quantum realm, coherences are needed. Perhaps they could be included by means of suitable wavepackets 
combined with the classical dynamics plus noise. This idea is still under investigation and will be the 
subject of future works.

\section*{Acknowledgments}

Support from CONICET is gratefully acknowledged.

\vspace{3pc}


\end{document}